\documentstyle[prl,aps,epsf]{revtex}
\def\<{\langle}
\def\>{\rangle}
\def\d{\partial}
\def\+{\dagger}

\firstfigfalse

\begin{document}
\twocolumn[\hsize\textwidth\columnwidth\hsize\csname@twocolumnfalse\endcsname
\preprint{CU-TP-xxx}
\title{QCD at Finite Isospin Density}
\author{D.T.~Son$^{1,3}$ and M.A.~Stephanov$^{2,3}$}
\address{$^1$ Physics Department, Columbia University, New York, New York 
10027}
\address{$^2$ Department of Physics, University of Illinois, Chicago, 
Illinois 60607-7059}
\address{$^3$ RIKEN-BNL Research Center, Brookhaven National Laboratory,
Upton, New York 11973}
\date{May 2000}
\maketitle

\begin{abstract}
QCD at finite isospin chemical potential $\mu_I$
has no fermion sign problem  and can be studied on the
lattice.  We solve this theory analytically in two limits:
at low $\mu_I$, where chiral perturbation
theory is applicable, and at asymptotically high $\mu_I$, where
perturbative QCD works.  At low isospin density the ground state is a
pion condensate, whereas at high density it is a Fermi liquid with
Cooper pairing.  The pairs carry the same quantum numbers as the
pion. This leads us to conjecture that the transition from 
hadron to quark matter
is smooth, which passes several tests. Our results imply
a nontrivial phase diagram in the space of temperature
and chemical potentials of isospin and baryon number.
\end{abstract}
%\pacs{12.38.Aw, 03.75.Fi, 11.30.Rd, 12.38.Mh}
\vskip 2pc]

{\em Introduction}.---Ample knowledge of QCD in the regime of finite
temperature and baryon density is crucial for understanding a wide
range of phenomena from 
heavy ion collisions to neutron stars and
cosmology.  First-principles lattice numerical Monte Carlo 
calculations provide a solid basis for our knowledge of the 
finite-temperature regime.  However, the regime of finite baryon chemical
potential $\mu_B$ is still inaccessible by Monte Carlo
because present methods of evaluating the QCD partition function
require taking a path integral with a measure which includes a {\em
complex} fermion determinant.  Ignoring the determinant (as in the popular
quenched approximation) leads to qualitatively wrong answers for
finite $\mu_B$~\cite{quenched}.  
Such a contrast to the case of $\mu_B=0$, where the
quenched approximation proved useful, comes from the fact
that the latter corresponds to an unphysical theory with pairs of
quarks of opposite baryon charges (conjugate quarks)
\cite{conjquarks}.
%A class of theories with similar
%physical properties is QCD with 2 colors of quarks, instead of
%3. Quarks in 2 color QCD are self-conjugate, which ensures positivity
%and applicability of lattice Monte Carlo methods
%\cite{2color-lattice}.
%Another such class are theories
%with quarks in adjoint color representations~\cite{2color}. 
This is one of the main reasons why our understanding of QCD at finite
baryon density is still rudimentary.  Many interesting phenomena,
such as color superconductivity and color-flavor locking
\cite{colorSC}, occur at finite
baryon density, beyond the reach of current lattice
techniques.  

To understand the regime of finite baryon density one would need
to follow the transition from hadronic to quark degrees of freedom 
by increasing the density of a {\em conserved} charge (such as baryon
number), i.e., without invoking the temperature.
This is the motivation for us to turn to QCD at
finite chemical potential $\mu_I$ of {\em isospin\/} (more precisely,
of the third component, $I_3$).
% $u$ and $d$ quarks are mutually conjugate.
Nature provides us with nonzero $\mu_I$ systems  in
the form of isospin-asymmetric matter.  These always contain both
isospin density and baryon density.  In any realistic setting
$\mu_I \ll \mu_B$.  In this paper, however, we shall consider
an idealization in which $\mu_I$ is nonzero while $\mu_B=0$.
Such a
system is unstable with respect to weak decays which do
not conserve isospin.  However, since we are interested in the
dynamics of strong interaction alone, one can imagine that all
relatively slow
electroweak effects are turned off.  Once this is done, we have a
nontrivial regime which, as has been emphasized recently in
\cite{AKW}, is
% has the advantage of being 
accessible by present lattice Monte Carlo methods, while being, as we shall
see, analytically tractable
in various interesting limits. As a result, the system we consider has 
a potential to 
improve substantially our understanding of cold dense QCD.  This regime carries
many attractive traits of two-color QCD~\cite{2color-lattice,2color}, 
but is realized
in a physically relevant theory --- QCD with three colors.

{\em Positivity and QCD inequalities}.---Since the Euclidean version
of our theory has a real and positive fermion determinant, some rigorous
results on the low-energy behavior can be obtained from QCD
inequalities \cite{ineqs,2color}.  In vacuum QCD, the latter rely on the
following property of the Euclidean Dirac operator ${\cal D} =
\gamma\cdot (\d + iA) + m$:
\begin{equation}
  \gamma_5 {\cal D} \gamma_5 = {\cal D}^\+ . \label{5d5}
\end{equation}
which, in particular, implies positivity $\det {\cal D}\ge0$.  
For the correlator of a generic meson $M=\bar\psi\Gamma\psi$, we can
write, using (\ref{5d5}) and the Schwartz inequality:
% Should it be Schwartz-Bunyakovsky?
\begin{eqnarray}
  & & \< M(x) M^\+(0) \>_{\psi,A} =
  -\< {\rm Tr} {\cal S}(x,0)\Gamma {\cal S}(0,x) \overline{\Gamma}\>_{A}=
  \nonumber\\
  & & \< {\rm Tr} {\cal S}(x,0) \Gamma i\gamma_5 
  {\cal S}^\+(x,0) i\gamma_5\overline{\Gamma} \>_{A}
  \le \< {\rm Tr} {\cal S}(x,0) {\cal S}^\+(x,0) \>_{A},
\label{corr}
\end{eqnarray}
where ${\cal S}\equiv {\cal D}^{-1}$ and
$\overline{\Gamma}\equiv\gamma_0\Gamma^\+\gamma_0$. The inequality is
saturated for mesons with $\Gamma=i\gamma_5\tau_i$, since ${\cal D}$
commutes with isospin $\tau_i$, which means that the pseudoscalar
correlators are larger, point-by-point, 
than all other $I=1$ meson correlators \cite{note1}.
As a consequence, one obtains an important restriction on the pattern
of the symmetry breaking: for example, it cannot be driven by a
condensate of $\<\bar \psi\gamma_5\psi\>$, which would give $0^+$
Goldstones.

At finite isospin density, $\mu_I\neq0$, positivity still holds
\cite{AKW} and certain inequalities can be derived (in contrast with
the case of $\mu_B\ne0$ when there is no
positivity).
%and hence no inequality applies
Now ${\cal D} = \gamma\cdot (\d + iA) + {1\over2}\mu_I\gamma_0\tau_3 +
m$, and Eq.\ (\ref{5d5}) is not true anymore, since the operation on
the right-hand side of (\ref{5d5}) changes the relative sign of $\mu_I$.
% as it does with the sign of $\mu_B$. 
However, provided $m_u=m_d$, 
interchanging up and down quarks compensates for this sign change
(the $u$ and $d$ quarks play the role of mutually
conjugate quarks \cite{conjquarks}), %in the sense of Ref.\cite{conjquarks} 
%with respect to
%$\mu_I$]
i.e,
\begin{equation}
  \tau_1 \gamma_5 {\cal D} \gamma_5 \tau_1 = {\cal D}^\dagger.
\label{t5d5t}
\end{equation}
Instead of isospin $\tau_1$ in (\ref{t5d5t}) one can also use $\tau_2$
(but not $\tau_3$).  Equation (\ref{t5d5t}), in place of the now invalid
Eq.\ (\ref{5d5}), ensures that $\det{\cal D}\ge0$. Repeating the derivation
of the QCD inequalities, by using (\ref{t5d5t}) we find that the lightest
meson, or the condensate, must be in channels $\bar\psi
i\gamma_5\tau_{1,2}\psi$, i.e., a linear combination of $\pi^-\sim\bar
u\gamma_5 d$ and $\pi^+\sim\bar d\gamma_5 u$ states.  Indeed, as shown
below, in the two analytically tractable regimes of small and large
$\mu_I$ the lightest mode is a massless Goldstone
% associated with spontaneously broken symmetry generated by $I_3$,
which is a linear combination of $\bar u\gamma_5 d$ and $\bar
d\gamma_5 u$.
% This is consitent with the inequalities.

{\em Small isospin densities}.---When $\mu_I$ is small compared to the
chiral scale (taken here to be $m_\rho$), we can use chiral
perturbation theory.  For zero quark mass and zero $\mu_I$ the pions
are massless Goldstones of the spontaneously broken
SU(2)$_L\times$SU(2)$_R$ chiral symmetry.  If the quarks have small
equal masses, the symmetry is only SU(2)$_{L+R}$.  The low-energy
dynamics is governed by the familiar chiral Lagrangian for the
% unitary matrix of 
pion field $\Sigma\in$ SU(2): ${\cal L} = {1\over4}f_\pi^2{\rm
Tr}[\d_\mu\Sigma\d_\mu\Sigma^\+ - 2m_\pi^2 {\rm Re}\Sigma]$, which
contains the pion decay constant $f_\pi$ and vacuum pion mass
$m_\pi$ as phenomenological parameters.
% [We will assume $m_\pi\ll m_\rho$, so
%there is a nontrivial range of $\mu_I$ and pion momenta 
%where the chiral theory is reliable.]
The isospin chemical potential further breaks SU(2)$_{L+R}$ down to
U(1)$_{L+R}$.  
Its effect can be included to leading order in $\mu_I$
without additional phenomenological parameters by promoting
SU(2)$_L\times$SU(2)$_R$ to a local gauge symmetry and viewing $\mu_I$
as the zeroth component of a gauge potential \cite{2color}.  Gauge
invariance thus fixes the way $\mu_I$ enters the chiral Lagrangian:
% The Lagrangian,
% describing the pion fields to lowest order in $\mu_I$, $m$ and
% derivatives has the form:
\begin{equation}
  {\cal L}_{\rm eff}= 
  {f_\pi^2\over4} {\rm Tr} \nabla_\nu \Sigma\nabla_\nu \Sigma^\+
%  - m_\pi\<\bar\psi\psi\>_{\rm vac} 
  - {m_\pi^2 f_\pi^2\over2}
  {\rm Re }{\rm Tr} \Sigma,
\label{leff}
\end{equation}
% where $\Sigma$ is an $SU(2)$ matrix whose slow phase 
% rotations according to $\Sigma\to U_L\Sigma U_R^\dagger$
% describe Goldstone modes. 
where the covariant derivative is defined as
\begin{equation}
  \nabla_0 \Sigma = \partial_0 \Sigma - {\mu_I\over2}
  (\tau_3\Sigma - \Sigma\tau_3).
%  ( I_3\Sigma - \Sigma I_3),
\end{equation}
% where $I_3$ is the generator of the third component of isospin of a
% quark field $I_3 = {\rm diag}(1/2,-1/2)$.
% \begin{equation}
% I_3 = \mat +1/2&0\\0&-1/2\emat.
% \end{equation}

By using (\ref{leff}), it is straightforward to determine vacuum
alignment of $\Sigma$ as a function of $\mu_I$ and the spectrum of
excitations around the vacuum.  We are interested in negative
$\mu_I$, which favors neutrons over protons, as in neutron stars.  The
results are very similar to two-color QCD at finite baryon density
\cite{2color}:

(i) For $|\mu_I|<m_\pi$, the system is in the same ground state as
at $\mu_I=0$: $\overline\Sigma=1$.  
This is because the lowest lying pion state costs a
positive energy $m_\pi-|\mu_I|$ to excite, which is impossible at zero
temperature.

%This is the normal vacuum of QCD. The excitations are pions
%with the masses given by the GOR relation: 
%$m_\pi^2 = {2m\langle\bar\psi\psi\rangle/ F_\pi^2}$.
%The energy of the pions at rest is shifted by $-\mu_I I_3$.
%The isospin density is zero in this case. This is the normal phase.

(ii) When $|\mu_I|$ exceeds $m_\pi$ 
% (i.e., $m_\pi$ divided by $|I_3|$ for pions) 
it is favorable to excite $\pi^-$ quanta, which form a Bose
condensate.  In the language of the effective theory, such a pion
condensate is described by a tilt of the chiral condensate 
$\overline\Sigma$,
% a second order phase
% transition occurs and the
% charged pions condense: $\pi^-$ for $\mu_I<-m_\pi$ and $\pi^+$
% for $\mu_I>m_\pi$. The condensate starts rotating as a function of
% $\mu_I/m_\pi$:
\begin{eqnarray}
  \overline\Sigma &=& \cos\alpha+i(\tau_1\cos\phi + \tau_2\sin\phi)
  \sin\alpha \, ,
\nonumber\\&&
\cos\alpha={m_\pi^2/ \mu_I^2} \, .
\label{alpha}
\end{eqnarray}
The tilt angle $\alpha$ is determined by minimizing the vacuum energy.
  The energy is
degenerate with respect to the angle $\phi$, corresponding to the
spontaneous breaking of the U(1)$_{L+R}$ symmetry  generated by $I_3$
in the Lagrangian
(\ref{leff}).  The ground state is a pion superfluid, with
one massless Goldstone mode.  Since we start from a theory with three
pions, there are two massive modes which can be identified with
$\pi_0$ and a linear combination of $\pi^+$ and $\pi^-$.  At the
condensation threshold, $m_{\pi_0}=m_\pi$ and the mass of the other
mode is $2m_\pi$, while for $|\mu_I|\gg m_\pi$ both masses approach
$|\mu_I|$.

The isospin density is found by differentiating the ground state
energy with respect to
% $-\partial {\cal L}_{\rm eff}/\partial
$\mu_I$ and is equal to:
\begin{equation}
  n_I = 
% -\partial {\cal L}_{\rm eff}/\partial\mu_I =
  f_\pi^2 \mu_I \sin^2\alpha
  =f_\pi^2 \mu_I\left(1-{m_\pi^4\over\mu_I^4}\right)
  , \quad |\mu_I| > m_\pi \, .
\label{ni}
\end{equation}
For $|\mu_I|$ just above the condensation threshold, $|\mu_I|-m_\pi\ll
m_\pi$, Eq.\ (\ref{ni}) reproduces the equation of state of the dilute
nonrelativistic pion gas \cite{2color}.
% The isospin
% density (\ref{ni}) is then approximately linear in $|\mu_I|-m_\pi$,
% which is in agreement with the well-known equation of state
% of a dilute Bose gas (see ref.\cite{...} for details).
%At $|\mu_I|\gg m_\pi$ Eq.\ (\ref{ni}) reduces to $n_I=f_\pi^2\mu_I$.

It is also possible to find baryon masses, i.e., the energy cost of
introducing a single baryon into the system.  The most interesting
baryons are those with lowest energy and highest isospin, i.e.\
neutron $n$ and $\Delta^-$ isobar.  There are two effects of $\mu_I$
on the baryon masses.  The first comes from the isospin of the
baryons, which effectively reduces the neutron mass by
${1\over2}|\mu_I|$ and the $\Delta^-$ mass by ${3\over2}|\mu_I|$.
Alone, this effect would lead to the formation of
baryon/antibaryon Fermi surfaces, manifested in nonvanishing
zero-temperature baryon susceptibility $\chi_B\equiv\partial
n_B/\partial \mu_B$ when $\mu_I>{2\over3} m_\Delta$.  However, long
before that, another effect turns on: the $\pi^-$'s in the condensate
tend to repel the baryons, lifting up their masses.  These effects
can be treated in the framework of the baryon chiral perturbation
theory \cite{Georgi}, giving
\begin{equation}
  m_n = m_N - {|\mu_I|\over2}\cos\alpha,\quad
  m_{\Delta^-} = m_\Delta - {3|\mu_I|\over2}\cos\alpha
\label{n,delta}
\end{equation}
in the approximation of nonrelativistic baryons.  
Equation (\ref{n,delta}) can be interpreted as follows: as a result of
the rotation (\ref{alpha}) of the chiral condensate, the 
% lowest lying
nucleon mass eigenstate becomes a superposition of vacuum $n$ and $p$
states.  The expectation value of the isospin in this state is
proportional to $\cos\alpha$ appearing in (\ref{n,delta}).  With 
$\cos\alpha$ given in Eq.(\ref{alpha}), we see that the two
mentioned effects cancel each other when $m_\pi\ll |\mu_I| \ll
m_\rho$.  Thus the baryon mass never drops to zero, and $\chi_B= 0$
%the baryon susceptibility vanishes 
at zero temperature in the region of applicability of the chiral
Lagrangian.

As one forces more pions into the condensate, the pions are packed
closer and their interaction becomes stronger.  When $\mu_I \sim
m_\rho$,
%the previous treatment based on 
the chiral perturbation theory breaks down.  To find the equation of
state in this regime, full QCD has to be employed.  As we have seen,
this can be done using present lattice techniques since the fermion
sign problem is not present at finite $\mu_I$, similar to the
two-color QCD \cite{2color-lattice}.

{\em Asymptotically high isospin densities}.---In the opposite limit
of very large isospin densities, or $|\mu_I|\gg m_\rho$, the description
in terms of quark degrees of freedom
applies since the latter are weakly interacting due to asymptotic
freedom.
%At large
%positive $\mu_I$, or $n_I$, 
%our system consists of $u$ quarks and $\bar{d}$ antiquarks
%(or, in the case of large negative $\mu_I$, $\bar{u}$ and $d$) which,
In our case of large negative $\mu_I$, or $n_I$, the ground state
consists of $d$ quarks and $\bar{u}$ antiquarks which, neglecting the
interaction, fill two Fermi spheres with equal radii $|\mu_I|/2$.
Turning on the interaction between the fermions leads to the
instability with respect to the formation and condensation of Cooper
pairs, similar, to some extent, to the diquark pairing at high baryon
density \cite{colorSC}.  In our case, $\mu_I<0$, the Cooper pair
consists of a $\bar u$ and a $d$ in the color singlet channel.  The
order parameter has the same quantum numbers as the pion condensate at
lower densities,
\begin{equation}
  \langle \bar{u} \gamma^5 d \rangle \ne 0.
  \label{ubard_cond}
\end{equation}
Because of Cooper pairing, the fermion spectrum acquires a gap $\Delta$ at
the Fermi surface, where  
\begin{equation}
  \Delta = 
%  \exp\biggl(-{\pi^2+4\over16}\biggr) 256 \pi^4
  b |\mu_I| g^{-5} e^{-c/g}, \qquad c = 3\pi^2/2
  \label{gap}
\end{equation}
where $g$ should be evaluated at the scale $|\mu_I|$.  This behavior
comes from the long-range magnetic interaction, as in the
superconducting gap at large $\mu_B$ \cite{Son:1999uk}.  The constant
$c$ is smaller by a factor of $\sqrt{2}$ compared to the latter case
due to the stronger one-gluon attraction in the singlet $q\bar{q}$
channel compared to the $\bar{\bf 3}$ diquark channel.  Consequently,
the gap (\ref{gap}) is exponentially larger than the diquark gap at
comparable baryon chemical potentials.  By using the methods of
\cite{Rockefeller}, one can estimate $b\approx 10^4$.

The perturbative one-gluon exchange responsible for pairing at large
$\mu_I$ does not distinguish $\bar u d$ and $\bar u \gamma_5 d$
channels: the attraction is the same in both. The $\bar u \gamma_5 d$
channel is favored by the instanton-induced interactions, which
explains the fact that the condensate is a pseudoscalar and breaks
parity.  This is consistent with our observation 
%mentioned above 
that QCD
inequalities also constrain the $I=1$ condensate to be a pseudoscalar
at any $\mu_I$.

{\em Quark-hadron continuity}.---Since the order parameter
(\ref{ubard_cond}) has the same quantum numbers and breaks the same
symmetry as the pion condensate in the low-density regime, it is
plausible that there is no phase transition along the $\mu_I$ axis.
In this case the Bose condensate of weakly interacting pions smoothly
transforms into the superfluid state of $\bar{u}d$ Cooper pairs.  The
situation is very similar to that of strongly coupled superconductors
with a ``pseudogap'' \cite{pseudogap}, and possibly of
high-temperature superconductors \cite{Randeria}.  This also parallels
the continuity between nuclear and quark matter in three-flavor QCD as
conjectured by Sch\"afer and Wilczek \cite{continuity}.  We hence
conjecture that, in two-flavor QCD, one can move continuously from the
hadron phase to the quark phase without encountering a phase
transition.  Since a first order deconfinement phase transition at
intermediate isospin chemical potential cannot be rigorously ruled out
(though it is unlikely, see below), this conjecture needs to be verified
by lattice calculations.
 
A number of nontrivial arguments support the continuity
hypothesis.  One notices that all fermions have a gap at large
$|\mu_I|$, which implies that 
%the baryon number susceptibility
$\chi_B=0$ at $T=0$.  This is also true at small $\mu_I$.  It is thus
natural to expect that $\chi_B$ remains zero at $T=0$ for all $\mu_I$,
which also suggests one way to check the continuity on the lattice.

Another argument comes from considering the limit of a large number of
colors $N_c$.  In finite-temperature QCD, the fact that the number of
gluon degrees of freedom is ${\cal O}(N_c^2)$ while that of 
% low lying
hadrons is ${\cal O}(N_c^0)$ hints at a first order
confinement-deconfinement phase transition.
At very large $\mu_I$ thermodynamic quantities such as the density of
isospin $n_I$ are proportional to $N_c$.  On the other hand, 
in the large $N_c$ limit 
the pion decay constant scales as $f_\pi^2={\cal
O}(N_c)$, and according to Eq.\ (\ref{ni}) the isospin density in the
pion gas is also proportional to $N_c$.  Physically, the repulsion
between pions becomes weaker as one goes to large $N_c$, thus more
pions are stacked at a given chemical potential.  As a result, the
$N_c$ dependence of thermodynamic quantities is the same in the quark
and the hadronic regimes.
%, although for seemingly different reasons.

%Another interesting argument in favor of the continuity%
%\footnote{Thanks to Larry McLerran for pointing this out.} 
%is the large $N_c$ matching of the two regimes. The density of the
%isospin $n_I$ in the large $\mu_I$ limit is given by the
%volume of the quark Fermi sphere 
%$n_I=(N_c/3)\mu^3/(8\pi^2)$, while at small $\mu_I$ it is 
%given by the pion gas formula (\ref{ni}). In the large $N_c$
%limit $F_\pi^2={\cal O}(N_c)$, and thus the two regimes match.
%The matching happens at a scale of order ${\cal O}(N_c^0\,\Lambda_{\rm
%QCD})$, the deconfinement scale.

{\em Confinement}.---
% Naively, one would think that if $\bar{u}$ and $d$
% quarks are packed at a very high density, the system becomes
% deconfined, signalling this by the existence of finite-energy excitations
% carrying fractional baryonic charge.  The obvious candidates for the
% latter are Landau quasiparticles near the quark Fermi surface.
% The opening of a BCS gap make the energy of these excitations larger
% than $\Delta$, but still finite.
At large $\mu_I$, gluons softer than $\Delta$ are not screened 
by the Meissner or by the Debye effect \cite{Rischke}: the condensate does
not break gauge symmetry (in contrast to the color superconducting
condensate \cite{colorSC}) and there are no low-lying color
excitations to screen the electric field.
%Indeed, the Meissner effect is absent
%because the condensate does not break gauge symmetry (in contrast to
%color superconducting condensate \cite{colorSC}).  
%The Debye screening is
%also absent, because on scales softer
%than $\Delta$ there are no charge excitations%
% in the medium
%: Cooper
%pairs are neutral, while fermions are too heavy to be excited.  
Thus, the gluon sector below the $\Delta$ scale is described by pure
gluodynamics, {\em which is confining}.  This means there are no
quark excitations above the ground state: all particles and holes must
be confined in colorless objects, mesons and baryons, just like in
vacuum QCD.  If there is no transition along the $\mu_I$ axis, we
expect confinement 
at all values of $\mu_I$.  Since the
running strong coupling $\alpha_s$ at the scale of
$\Delta$ is small,
the confinement scale $\Lambda_{\rm QCD}'$ (which is, in general,
different from $\Lambda_{\rm QCD}$)
% perhaps larger than $\Lambda_{\rm QCD}$ since the running is faster
% in pure gluodynamics than in QCD with fermions
is much less than $\Delta$.  At large $\mu_I$, we thus predict a
temperature driven deconfinement phase transition at a temperature
$T_c'$ of order $\Lambda_{\rm
QCD}'$, which is expected to be of first order as in pure gluodynamics.
Since $\Lambda_{\rm QCD}'\ll\Delta$ the hadronic
spectrum is similar to that of a heavy quarkonium,
%spectrum follows the pattern of the constituent heavy
% quark model,
with $\Delta$ playing the role of the heavy quark mass.

\begin{figure}
%  \centerline{\psfig{file=tmui.eps,width=.4\textwidth}}
  \def\epsfsize #1#2{0.5#1}
  \epsfbox{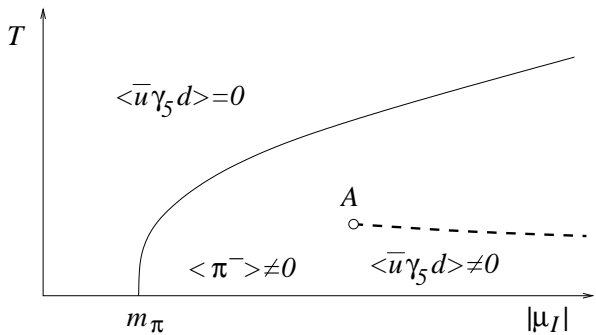}
  \caption[]{Phase diagram of QCD at finite isospin density.}
\label{fig:tmui}
\end{figure}

{\em The $(T, \mu_I)$ phase diagram}.---By considering nonzero
$\mu_I$, we make the phase diagram of QCD three dimensional:
$(T,\mu_B,\mu_I)$.
Two planes in this three-dimensonal space are of a special interest:
the $\mu_B=0$ $(T,\mu_I)$ plane, which is completely accessible
by present lattice techniques, and the $T=0$ $(\mu_I,\mu_B)$ plane, 
where the neutron star matter belongs.
% and where all possible critical points are {\em quantum}.
Two phenomena determine the $(T, \mu_I)$ phase plane 
(Fig.\ref{fig:tmui}):  pion condensation and
confinement.  
At sufficiently high temperature
the condensate (\ref{ubard_cond}) melts (solid line in
Fig.~\ref{fig:tmui}). For large $\mu_I$, this
critical temperature is proportional to the BCS
gap (\ref{gap}). There are two phases which differ by symmetry: the high
temperature phase, where the explicit flavor $U(1)_{L+R}$
symmetry is restored, and the low-temperature phase, where this symmetry
is spontaneously broken.  The phase
transition is in the O(2) universality class~\cite{noteGinzburg}.
 The critical temperature $T_c$ vanishes at
$\mu_I=m_\pi$ and is an increasing function of $\mu_I$ in both
regimes we studied:
$|\mu_I|\ll m_\rho$ and $|\mu_I|\gg\Lambda_{\rm QCD}$.  Thus, it is
likely that $T_c(\mu_I)$ is a monotonous function of $\mu_I$. 
In addition, at large
$\mu_I$, there is a first order 
%temperature driven
deconfinement phase transition at 
%a temperature
 $T_c'$ 
much lower than $T_c(\mu_I)$.  Since there is no 
%deconfiment 
phase transition at $\mu_I=0$ (for small $m_{u,d}$) or at $T=0$ (assuming
quark-hadron continuity), 
this first-order line must end at some point $A$ on the $(T, \mu_I)$
plane
(Fig.~\ref{fig:tmui}).
The exact location of $A$ should be determined by
lattice calculations; one of the possibilities is drawn in 
Fig.~\ref{fig:tmui}.

{\em The $(\mu_I, \mu_B)$ phase diagram}.---%
% The phase diagram in the
% $(\mu_I,\mu)$ plane  might be quite complicated.
This phase diagram deserves a separate study.  Here we shall only
consider the regime $|\mu_I|\gg\mu_B$ 
(the opposite limit $\mu_B\gg|\mu_I|$ was considered in Ref.\ 
\cite{cristalline}).
%When $\mu_B=0$ and
%$|\mu_I|\gg\Lambda_{\rm QCD}$
%is asymptotically large 
%we have seen that
%the system is a superfluid with a gap $\Delta$.  
Finite $\mu_B$
provides a mismatch between $\bar{u}$ and $d$ Fermi spheres,
which makes the
superconducting state unfavorable at some value of $\mu_B$ of
order $\Delta$.
%Increasing $\mu_B$ amounts to giving the $\bar{u}$ and the
%$d$ different chemical potentials.  It is clear that the
%superconducting state becomes unfavorable at some value of $\mu_B$ of
%order $\Delta$.  
It is known \cite{FFLO} that the destruction of this state
%the superfluid phase 
occurs through {\em two} phase transitions:
% At small $\mu_B$, the system is in the same state as at $\mu_B=0$, 
% since excitations carrying baryon number require finite energy.  
%As one increases $\mu_B$, 
one at $\mu_B$ slightly below $\Delta/\sqrt{2}$ and 
another at $\mu_B=0.754\Delta$.  The ground state between the two
phase transitions is the
Fulde-Ferrell-Larkin-Ovchinnikov (FFLO) state \cite{FFLO},
characterized by a spatially modulated superfluid order parameter
$\<\bar u\gamma_5 d\>$ with a wavenumber of order
$2\mu_B$.
% How the spatial dependence looks like is still not quite known,
% mostly because the FFLO state has not been observed in experiment.
% At the second order phase transition to the normal phase the 
% wavenumber should be approximately $2.4\mu_B$.
The FFLO state has the same symmetries as the inhomogeneous pion
condensation state which might form in electrically neutral nuclear
matter at high densities \cite{picond}.  It is conceivable that the
two phases are actually one, i.e., continuously connected on the
$(\mu_I,\mu_B)$ phase diagram.  
%The FFLO state persists only until
%$\mu_B=0.754\Delta$ when it goes through a second-order phase
%transition to a $\<\bar u\gamma_5 d\>=0$ state.  The latter is a
%$p$-wave 
%color superconductor with one-flavor diquark condensates
%$\<uu\>$ and $\<dd\>$, due to the attraction between quarks of each
%flavor.

% Here we shall not pursue further the details of the phase diagram in
% the $(\mu,\mu_I)$ plane, which
% seems to be interesting enough to 
% deserves a separate study.  We would like to make only two more
% remarks.  First, the FFLO state realized when
% $0.707<\mu_B/\Delta<0.754$ has the same symmetry as the inhomogeneous
% pion condensation state which might form in electrically neutral
% nuclear matter at high densities \cite{picond}.  It is conceivable
% that the two phases are actually one, i.e., continuously connected on
% the $(\mu,\mu_I)$ phase diagram.  Second, at each value of $\mu_I$
% there is a minimal $\mu_B$ (approximately $m_N$ at small $\mu_I$ and
% $\Delta/\sqrt{2}$ at large $\mu_I$), below which $n_B\equiv0$.  Thus
% there must be a line, running roughly parallel to the $\mu_I$ axis,
% separating regions with $n_B=0$ and $n_B\ne0$.  The line is first
% order at very small $\mu_I$ (since it separates vacuum from nuclear
% matter) and at asymptotically large $\mu_I$.  In between, there might
% be an interval where the transition is second or higher order.
% For example, we think that a small $\mu_I$ can unbind nuclear matter.

The authors thank L.~McLerran, J. Kogut, R.~Pisarski, and E.~Shuryak for
discussions, the DOE Institute for Nuclear Theory
 at the University of Washington for
its hospitality, and K.~Rajagopal for drawing their attention to
Ref.\ \cite{FFLO}.


\begin{thebibliography}{99}
\vspace{-0.5cm}
\bibitem{quenched}
                I. Barbour {\em et al},
		%N.-E. Behilil, E. Dagotto, F. Karsch, 
                %A. Moreo, M. Stone, and H.W. Wyld, 
                Nucl. Phys. {\bf B275} [FS17], 296 (1986);
                J.B. Kogut, M. P. Lombardo and D.K. Sinclair,
                Phys. Rev. D {\bf 51}, 1282 (1995); 
                Nucl. Phys. B, Proc. Suppl. {\bf 42}, 514 (1995).
 

\bibitem{conjquarks}
%\bibitem{Stephanov:1996ki}
M.A.~Stephanov,
%``Random matrix model of QCD at finite density and the nature of the 
%quenched limit,''
Phys.\ Rev.\ Lett.\ {\bf 76}, 4472 (1996).
% [hep-lat/9604003].

\bibitem{colorSC}
	B.C. Barrois, 
%{\em Non-perturbative effects in dense quark matter}, 
        PhD Thesis, California Institute of Technology, 1979;
        D. Bailin and A. Love, Phys. Rep. {\bf 107}, 325 (1984);
        M. Alford, K. Rajagopal and F. Wilczek,
                Phys. Lett. {\bf B422}, 247 (1998),
                Nucl. Phys. {\bf B537}, 443 (1999);
        R. Rapp, T. Sch\"afer, E.V. Shuryak and M. Velkovsky,
                Phys. Rev. Lett. {\bf 81}, 53 (1998); 
		Ann. Phys. (N.Y.) {\bf 280}, 35 (2000).

\bibitem{AKW}
	M.~Alford, A.~Kapustin and F.~Wilczek,
	%``Imaginary chemical potential and finite 
	%fermion density on the lattice,''
	Phys.\ Rev.\  D {\bf 59}, 054502 (1999).
	%[hep-lat/9807039].

\bibitem{2color-lattice}
        E. Dagotto, F. Karsch, and A. Moreo,
        Phys. Lett. {\bf  169B}, 421 (1986);
        E. Dagotto, A. Moreo, and U. Wolff, 
	Phys. Rev. Lett. {\bf 57}, 1292 (1986); 
	Phys. Lett. B {\bf  186}, 395 (1987);
%\bibitem{SU2lattice.new}
        S. Hands, J.B. Kogut, M.-P. Lombardo,
        S.E. Morrison, Nucl.\ Phys.\  {\bf B558}, 327 (1999);
        S. Hands and S.E. Morrison, hep-lat/9902012; hep-lat/9905021.

\bibitem{2color}
% \bibitem{Kogut:1999iv}
J.B.~Kogut, M.A.~Stephanov, and D.~Toublan,
% ``On two-color QCD with baryon chemical potential,''
Phys.\ Lett.\  B {\bf 464}, 183 (1999);
% [hep-ph/9906346]
% \bibitem{Kogut:2000ek} 
J.B.~Kogut, 
M.A.~Stephanov, D.~Toublan, J.J.~Verbaarschot and A.~Zhitnitsky,
%``QCD-like theories at finite baryon density,''
%hep-ph/0001171, to be published in Nucl. Phys. B.
Nucl. Phys. {\bf B582}, 477 (2000).

\bibitem{ineqs}
		D. Weingarten, Phys. Rev. Lett. {\bf 51}, 1830 (1983);
		E. Witten, Phys. Rev. Lett. {\bf 51}, 2351 (1983);
		S. Nussinov, Phys. Rev. Lett. {\bf 52}, 966 (1984);
		D. Espriu, M. Gross and J.F. Wheater, Phys. Lett.
		{\bf 146B}, 67 (1984).

\bibitem{note1} It is important, as is the case for $I=1$, that there
is no disconnected piece after $\psi$ integration in (\ref{corr}). The
proof does not apply, for example, to $\sigma$ or $\eta$ 
meson correlators, $\Gamma=1,\gamma_5$.

\bibitem{Georgi} See, e.g., H.~Georgi, {\em Weak Interaction and
Modern Particle Theory} (Benjamin-Cummings, Menlo Park, 1984).

\bibitem{Son:1999uk}
D.T.~Son,
%``Superconductivity by long-range color magnetic interaction in  
% high-density quark matter,''
Phys.\ Rev.\ D {\bf 59}, 094019 (1999).
% [hep-ph/9812287].

\bibitem{Rockefeller} W.E.~Brown, J.T.~Liu, and H.-C.~Ren,
	Phys.\ Rev.\ D {\bf 61}, 114012 (2000);
	%hep-ph/9908248, 
%	hep-ph/9912409, hep-ph/0003199.
        {\bf 62}, 054016 (2000); {\bf 62}, 054013 (2000).

\bibitem{pseudogap} A.J.~Leggett, J. de Phys. {\bf 41}, C7-19 (1980);
P.~Nozi\`eres and S.~Schmitt-Rink, J. Low Temp. Phys. {\bf 59}, 195
(1985).

\bibitem{Randeria} See, e.g., M.~Randeria, cond-mat/9710223 and
references therein.

\bibitem{continuity}
%\bibitem{Schafer:1999ef}
T.~Sch\"afer and F.~Wilczek,
%``Continuity of quark and hadron matter,''
Phys.\ Rev.\ Lett.\  {\bf 82}, 3956 (1999).
%[hep-ph/9811473].

\bibitem{Rischke} This is similar to the behavior of the unbroken
SU(2)$_c$ sector of two-flavor color superconductors (see
D.H.~Rischke, 
%nucl-th/0001040.
Phys. Rev. D {\bf 62}, 034007 (2000)).

\bibitem{noteGinzburg}
	The width of the Ginzburg region is suppressed
	by $(\Delta/\mu_I)^4$ at large $\mu_I$
	and also by $1/N_c^2$ at large $N_c$: %as at the QCD chiral
					 %transition.
	J.B.~Kogut, M.A.~Stephanov and C.G.~Strouthos,
	%``Critical region of the finite temperature chiral transition,''
	Phys.\ Rev.\ D {\bf 58}, 096001 (1998).
	%[hep-lat/9805023].

\bibitem{cristalline} M.~Alford, J.~Bowers and K.~Rajagopal,
hep-ph/0008208.

\bibitem{FFLO} P.~Fulde and A.~Ferrell, Phys. Rev. {\bf 135}, A550
(1964); A.I.~Larkin and Yu.N.~Ovchinnikov, Sov. Phys. JETP {\bf 20},
762 (1965);  see also Ref.\ \cite{cristalline}.

\bibitem{picond} A.B.~Migdal, 
	%Zh. Eksp. Teor. Fiz. {\bf 61}, 2210 (1971) 
	Sov. Phys. JETP {\bf 36}, 1052 (1973); 
	R.F.~Sawyer,
	Phys. Rev. Lett. {\bf 29}, 382 (1972); 
	D.J.~Scalapino, Phys. Rev. Lett. {\bf 29}, 386 (1972).
%See, e.g., G.~Baym and D.K.~Campbell in: {\em Mesons
%and Nuclei}, v. 3, eds. M.~Rho and D.~Wilkinson (North Holland, 1979).

\end{thebibliography}
\end{document}